\documentclass[12pt]{iopart}
\usepackage[mathscr]{eucal}
 \usepackage{setstack}
\begin{document}
\title{In-medium Yang-Mills equations: a derivation and canonical
quantization}
\author{M. K. Djongolov$^{1)}$, S.Pisov$^{2)}$, V.Rizov$^{2)}$\\
\itshape 1) Department of Physics and Astronomy, University of Tennessee,\\ 
 e-mail: martin@spock.phys.utk.edu\\
\itshape 2) University of Sofia, Faculty of Physics, \\
 e-mail: pisov@phys.uni-sofia.bg; rizov@phys.uni-sofia.bg}
\maketitle
\begin{abstract}
The equations for a Yang-Mills field in a medium are derived in 
the approximation of linear response to an external field. Introducing tensors
of generalized susceptibilities, the in-medium equations are written in a
form similar to the in-medium Maxwell equations. The non-Abelian character of 
the gauge group $G$ is reflected in the presence of susceptibility tensors
with no counterpart in electrodynamics and the necessity to define modified 
"electric" and "magnetic" fields (apart from the "color" induction vectors).
which enter the in-medium equations. The latter reduce to $dimG$ copies of
Maxwell in-medium equations in the approximation up to second order 
in the gauge coupling constant. For a medium uniformly moving, a canonical 
quantization is performed for a family of Fermi-like gauges in
the case of constant and diagonal (in the group indices) tensors
of "electric" permittivity and "magnetic" permeability. The physical
subspace is defined and the gauge field propagator is evaluated
for a particular choice of the gauge. It is applied firstly for evaluation 
of the cross-section of elastic quark scattering
in the Born approximation. Possible applications to Cherenkov-type
gluon radiation are commented briefly.
\end{abstract}
\section{Introduction}
In some developments of the Yang-Mills theories one arrives at
situations when the encountered states can be effectively 
described as media, possessing non-Abelian polarization properties.
We list a few of these examples for sake of brevity.

Among the first prototypes for a non-Abelian polarizable medium
one may take the QCD vacuum state in the aspect considered by
Savvidy \cite{1}, Fukuda and Kazama \cite{2}, Adler \cite{3},
Nielsen and Olesen \cite{4}. In \cite{5} and \cite{6} Baker and
Zachariasen used the Yang-Mills vacuum as a polarizable medium in
order to derive a confining potential. They also solved the Dyson
equations for the gluon propagator $\Delta_{\mu\nu}(q)$ for $q^2
\rightarrow 0$ evaluating the color susceptibility of the vacuum
in the weak field approximation. The polarization properties of
the non-Abelian $SU(2)$ vacuum were exploited in \cite{7,8} for an
approximate classical description of some non-perturbative effects
such as solutions corresponding to non-vanishing gauge field
"condensates".

Another and more recent set of examples comes from the
theoretical study of some not yet confirmed states of the hadron
matter - hadron matter at very high densities and quark-gluon
plasma (QGP) at very high temperatures where one expects the
quarks and gluons to be in a deconfined phase. In a systematic
study performed in the framework of kinetic theory the color
conductivity and other transport coefficients of non-Abelian
plasma were derived by Heinz in \cite{9,10} in the linear response
approximation. The response of QGP beyond the linear
approximation was considered in \cite{11} yielding a color
dielectric susceptibility modified by the non-linear corrections.
Typical non-Abelian medium effects on the color dielectric
constant $\epsilon$ were further investigated in \cite{12} also on
the base of kinetic theory.

The approach of Heinz has been generalized in \cite{13D,14D} to 
include the effects of fluctuations in deriving effective 
transport equations for QCD. A clear review of the methods of 
semi-classical transport theory for non-Abelian plasma can be found in
\cite{15D}.

From the analysis of the phenomenon of color superconductivity of
hadron matter at high density \cite{13,14}, it has been found 
that the medium surrounding the liberated quarks exhibits a
polarization behavior characterized by an effective color
dielectric constant $\epsilon \gg 1$, the magnetic permeability
remaining close to one. The estimate $\epsilon \gg 1$ has also been
obtained in \cite{15} evaluating the photon self-energy for a color
superconductor. The case of two-flavor color superconductor with 
prescribed dielectric susceptibility and magnetic permeability
of the medium was considered by D. Litim and C. Manuel \cite{19D}
applying the methods of transport theory and extending the results of 
Rischke, Son and Stephanov to finite temperature.

Finally, we mention models such as the chromodielectric model
proposed by Friedberg and Lee \cite{16,17} and developed in
\cite{18,19,20,21,26D,27D}. Aimed to describe an effective mechanism for
the confinement phenomena, this model exploits a dielectric
function depending on a classical field in order to account for
the interaction of the QCD matter. Also in \cite{28D,29D} Dremin introduced
a refraction coefficient $n$ phenomenologically for a hadron medium,
exploiting the relation between $n$ and the elastic forward scattering
amplitude of light derived in classical electrodynamics. Among other applications, 
for a quark-parton
medium he used the appearance of a so postulated $n$ to consider the 
emission of Cherenkov-type gluons (for $n>1$) in order to explain
the ring-like structures observed in hadron-hadron or central nucleus-nucleus collisions
as generated by gluon jets produced in such processes (see e.g.\cite{30D}).

Having in mind these examples, our view-point is that, if in the
interactions of a medium with Yang-Mills fields one encounters
manifestations of its polarization-type response to the gauge
fields, then it is worth developing an in-medium Yang-Mills
theory and apply it to situations where such a response is found
to be essential (like, e.g. the quark-gluon plasma).

In this paper we derive the classical Yang-Mills equations for a gauge group $G$ in a
medium in the case when the medium responses linearly to an external gauge field
(i.e. we derive the analogs of the
macroscopic Maxwell equations for a gauge theory) and perform a canonical
quantization of the arising theory. In the derivation of the
classical in-medium gauge field equations we follow the method of
Akhiezer and Peletminsky \cite{22} in their derivation of the
macroscopic Maxwell equations. In Section 2 we introduce some of
the notations and write down the equations for a system of
interacting spinor and gauge fields in the presence of a weak
external gauge field. In Section 3 these equations are averaged
by means of a non-equilibrium statistical operator in the linear
approximation with respect to the external field, and the averaged "color" electric and 
magnetic fields $(E^a,B^a)$ are introduced. There appear three types 
of (generalized) conductivity tensors 
defined in terms of the relevant retarded Green's functions. The first is a straightforward 
generalization to the Yang-Mills case of the "classical electrodynamic" conductivity tensor,
while the second and the third one appear due to the (eventually) non-commutative
character of the gauge group. In Section 4 the relevant polarization 
tensors, characterizing the medium, are defined, two of them having 
direct analogs in the classical case, 
and other two originating from the non-commutativity of the gauge group.
Assuming a linear relation between the polarization tensors and the fields $(E^a,B^a)$,
we define the (non-Abelian) susceptibility tensors.
In Section 5 we define the generalized electric and magnetic
inductions, the (generalized) magnetic permeability 
and electric permittivity tensors $\mu^{ab}$,$\epsilon^{ab}$,
and write down the derived in-medium Yang-Mills equations. As in the standard case, 
four of these equations describe the "color" induction vectors. In order to write
the remaining equations, we define modified "color" fields $(E'^a,B'^a)$
obtained from $(E^a,B^a)$ by means of the "non-Abelian" susceptibility tensors.
In Section 6 we perform an approximation up to second order in
the gauge coupling constant which simplifies the
in-medium equations significantly reducing them to a form resembling closely the
Maxwell in-medium equations.
Defining the constitutive equations for the
medium, we obtain relations between $\epsilon^{ab}$, $\mu^{ab}$,
and the conductivity tensors in the approximation considered.
These relations appear as natural generalizations of their
analogs in the classical macroscopic electrodynamics. 
We follow Brevik and Lautrup
\cite{23} and Nakanishi \cite{24,25} in their treatment of the
electromagnetic field which avoids work with constrained systems.
Although in \cite{23} one deals with in-medium electromagnetic
field, and \cite{24,25,26} consider the electromagnetic field in
vacuum, these papers use a common approach, namely a Lagrangian
formulation involving an auxiliary scalar field whose action
on the Fock space is designed to separate the physical subspace.
For a medium moving uniformly, Section 5 provides a Lagrangian 
formulation of the previously
obtained equations introducing a set of $dim (G)$ auxiliary fields
$B^a$. For sake of simplicity we consider constant and diagonal tensors 
$\epsilon$ and $\mu$. From the equal time commutation relations, 
in Section 6 we evaluate the commutators at arbitrary space-time points. 
In Section 7  the fields' Fourier transforms are introduced. 
In the resulting Fock space a physical subspace is defined by means of the 
operators $B^a$. Finally, for a particular choice of the gauge we evaluate
the gauge field propagator. As a first application the cross-section for
elastic quark-quark scattering is given in the Born approximation. 
It is noticed that the in-medium
Yang-Mills equations considered here provide a natural framework for
the Dremin's approach to explain the specific ring-like structures in hadron 
processes as resulting from a Cherenkov-type gluon radiation. Other applications,
like, e.g. color superconductivity, lie beyond the scope of this paper.

\section{Field equations}
Let $G$ be a compact $r$-dimensional simple matrix Lie group. Let
$T^a$ $(a = 1,..., r)$ be a basis of hermitian matrices in its
Lie algebra $\mathcal{G}$ with commutation relations
\[
[T^a, T^b] = {\it i} f^{abc}T^c,
\]
\noindent and normalized by
\[
tr (T^aT^b)=\frac{1}{2}\delta^{ab}.
\]
Lower case Latin letters $a,b,c,...$ are used for the group
/algebra indices and here and below summation over repeated
indices is understood.

Let $\hat{\Psi}(x)$ be a quantized spinor field with values in a
vector space $\mathcal{W}$, i.e. $\hat{\Psi}(x) \in
\hat{\mathcal{S}}\otimes\mathcal{W}$, where $\hat{\mathcal{S}}$
is the space of operator-valued spinors over Minkowski space $M$,
$x \in M$. We suppose that $\mathcal{W}$ carries a linear
representation $\omega \rightarrow R(\omega)$ of G, $\omega \in
G$, and that $\mathcal{W}$  is equipped with a scalar product
such that $R(\omega)$ is unitary.

Given a quantized Yang-Mills gauge potential
$\hat{\mathscr{A}}_{\mu}(x) =
\hat{A}^a_{\mu}(x)T^a$ $(\mu = 0,...,3)$, the gauge covariant derivative of the
spinor field $\hat{\Psi}(x)$ reads (in coordinates
$(x^{\mu})=(x^0,\vec{x})$)
\[ {\mathscr{D}_{\mu}(\hat{\mathscr{A}}) \hat{\Psi}(x) =
\partial_{\mu}\hat{\Psi}(x) + {\it i} g
R(\hat{\mathscr{A}}_{\mu}(x))\hat{\Psi}(x) =
\partial_{\mu}\hat{\Psi}(x) +
{\it i} g \hat{A}^a_{\mu}(x)R(T^a)\hat{\Psi}(x),} \]

\noindent where $\mathscr{D}_{\mu}(\hat{\mathscr{A}})$ denotes
covariant differentiation with respect to
$\hat{\mathscr{A}}_{\mu}$, and $g$ is a gauge coupling constant
($R(T^a)$ stands for the generators of G in the representation
$R$). We suppose, in addition, that the spinor field $\hat{\Psi}$
is put in an external (classical) gauge field $f_{\mu\nu}$ with
gauge potential $a_{\mu}= a^c_{\mu}T^c$, i.e.
\[
f_{\mu\nu} = \partial_{\mu}a_{\nu} - \partial_{\nu}a_{\mu} + {\it i} g
[a_{\mu},a_{\nu}],
\]
\noindent and $f_{\mu\nu}$ is determined by an external current
\[
j^{\mu} = j^{a,\mu}T^a,
\]
\noindent so that
\begin{equation}
\label{1.1}
\mathscr{D}_{\mu}(a)f^{\mu\nu} =  \partial_{\mu}f^{\mu\nu} +
{\it i} g [a_{\mu},f^{\mu\nu}]=g j^{\nu}.
\end{equation}

Fixing the units such that $\hbar=c=1$, the Minkowski metric
$(g_{\mu\nu})$ = $diag(1, -1, -1,- 1)$, and denoting the operators
$(\partial_n)$ = $(-\partial^n)$ = $\vec{\nabla}$, we also introduce the
components of the analogs of the electric and magnetic fields
\begin{eqnarray}
\hat{E}^{a,k} & = & -\hat{\mathscr{F}}^{a,0k}, \nonumber \\
\\
\hat{B}^{a,k} & = &-\frac{1}{2}\epsilon^{kmj}\hat{\mathscr{F}}^{a,mj},
\nonumber
\end{eqnarray}
\noindent where $a=1,...,r; k=1,2,3 , \epsilon^{123}=1$ and
\[
\hat{\mathscr{F}}^{\mu\nu}=\partial^{\mu}\hat{\mathscr{A}}^{\nu} -
\partial^{\nu}\hat{\mathscr{A}}^{\mu}
+ {\it i} g [\hat{\mathscr{A}}^{\mu},\hat{\mathscr{A}}^{\nu}] =
\hat{F}^{a,\mu\nu}T^a.
\]

The gauge potentials $\hat{\mathscr{A}}_{\mu}$ are taken in the temporal
gauge, $\hat{\mathscr{A}}_0=0$, satisfying
$\partial_k\hat{\mathscr{A}}_k=0$. We fix also
$\partial_{\mu}a^{\mu}=0$. 
The equations for $\hat{\mathscr{A}}(x)$ are
\begin{eqnarray}
\label{1.2}
\partial_0\hat{E}^{a,k}(x) = 
-\epsilon^{kmj}(\mathscr{D}^m(\hat{\mathscr{A}})\hat{B}^j(x))^a -
g(\hat{I}^{a,k}(x)+j^{a,k}(x^0,\vec{x})),\\
\label{1.3}
\partial_0\hat{B}^{a,k}(x) = 
\epsilon^{kmj}(\mathscr{D}^m(\hat{\mathscr{A}})\hat{E}^j(x))^a, \\
\label{1.4}
(\mathscr{D}^k(\hat{\mathscr{A}})\hat{B}^k(x))^a = 0, \\
\label{1.5}
(\mathscr{D}^k(\hat{\mathscr{A}})\hat{E}^k(x))^a  = g
(\hat{I}^{a,0}(x) + j^{a,0}(x^0,\vec{x})).
\end{eqnarray}
\noindent Here $\hat{I}_{\nu}(x)=\hat{I}^a_{\nu}(x)T^a$ is the
spinor current with components
\begin{equation}
\label{1.7}
\hat{I}^a_{\nu}(x) =
\hat{\Psi}^{\dag}(x)\gamma_{\nu}R(T^a)\hat{\Psi}(x).
\end{equation}
The equation for $\hat{\Psi}(x)$ reads
\begin{equation}
\label{eqn_of_motion}
\fl {\it i}\gamma^0\partial_0\hat{\Psi}(x) + {\it i}
\gamma^n\mathscr{D}_n(\hat{\mathscr{A}})\hat{\Psi}(x) + {\it i}
\gamma^0\mathscr{D}_0(a)\hat{\Psi}(x) + {\it i}
\gamma^k\mathscr{D}_k(a)\hat{\Psi}(x) - m\hat{\Psi}(x) = 0,
\end{equation}
\noindent where $\gamma^{\lambda}\gamma^{\mu} +
\gamma^{\mu}\gamma^{\lambda} = 2g^{\lambda\mu}{\bf 1}$. It follows from
equation (\ref{eqn_of_motion}) that
\begin{equation*}
\label{1.9}
\mathscr{D}^{\mu}(\hat{\mathscr{A}})\hat{I}_{\mu} = - {\it i} g
[a^{\mu},\hat{I}_{\mu}].
\end{equation*}
In absence of the exterior gauge field the energy functional has the form
\[
\hat{H_0} = \int dx^3\{\frac{1}{2}\sum^r_{a=1}(\vec{E}^a\cdot\vec{E}^a +
\vec{B}^a\cdot\vec{B}^a) + \hat{\Psi}^\dag\gamma^0(
-\it i\gamma^k\mathscr{D}_k(\hat{\mathscr{A}})\hat{\Psi}+m\hat{\Psi})\}.
\]
\section{Response of the system to the exterior gauge field}
\subsection{Density operator and averaged field equations}
We suppose that at infinite past $(x^0=-\infty)$ in the absence of the
exterior field the system of interacting fermion and gauge fields is in
equilibrium at temperature $1/\beta$, described by the density
operator $\hat{\rho_0}=exp(-\beta\hat{H})$. At time $x^0=-\infty$ the
exterior field $a^\mu$ is switched in adiabatically and the system is
described by an non-equilibrium density operator $\hat{\rho}(x^0)$ satisfying
\begin{equation}
\it i\partial_0\hat{\rho}(x^0)=[\hat{H}(x^0),\hat{\rho}(x^0)],
\end{equation}
\noindent with the initial condition $\hat{\rho}(-\infty)=\hat{\rho}_0$. The total
Hamiltonian is
\begin{equation}
\hat{H}(x^0)=\hat{H_0}+g\hat{V}(x^0),
\end{equation}
\noindent where $\hat{V}(x^0)$ describes the interaction of the fermions
with the gauge potential $a^c_\mu$,
\begin{equation}
\hat{V}(x^0) = \int d^3x\hat{I}^{c,\mu}(\vec{x})a^c_\mu(x^0,\vec{x}).
\end{equation}
It is convenient to go to the interaction (Dirac) representation, defined by
the splitting of the Hamiltonian 
\[
\hat{H}=\hat{H}(x^0)=\hat{H_0}+g\hat{V}(x^0).
\]%for each operator $
\noindent In this representation the density operator
$\hat{\rho}_{\mathcal{D}}(x^0)$ satisfies
\begin{equation}
\label{2.4}
\it i\partial\hat{\rho}_{\mathcal{D}}(x^0) =
g[\hat{V}_{\mathcal{D}}(x^0),\hat{\rho}_{\mathcal{D}}(x^0)],
\end{equation}
\noindent and the initial condition
$\hat{\rho}_{\mathcal{D}}(-\infty)=\hat{\rho}_0$, where
\begin{equation}
\label{2.5}
\hat{V}_{\mathcal{D}}(x^0) = \int
d^3x\hat{J}^\mu(x) a_\mu(x),
\end{equation}
\noindent and $\hat{J}^\mu$ is the fermion current in the Dirac
representation
\begin{equation}
\label{2.6}
\hat{J}^\mu(x) \equiv \hat{I}^\mu_{\mathcal{D}}(x) = exp(\it i x^0
\hat{H_0})\hat{I}^\mu(x)exp(-\it i x^0 \hat{H_0}).
\end{equation}
\noindent Combining equation (\ref{2.4}) and the initial condition in the
integral equation
\[
\hat{\rho}_\mathcal{D}(x^0) = \hat{\rho}_0 - \it i g \int_{-\infty}^{x^0}
dt [\hat{V}_\mathcal{D}(t),\rho_\mathcal{D}(t)],
\]
after the first iteration one finds the well known representation
\begin{equation*}
\label{2.7}
\hat{\rho}_\mathcal{D}(x^0) = \hat{\rho}_0 - \it i g
\int_{-\infty}^{+\infty} dt
\theta(x^0-t)[\hat{V}_\mathcal{D}(t),\hat{\rho}_0].
\end{equation*}
\noindent In other terms, the linear approximation of $\hat{\rho}_\mathcal{D}$ with
respect to the exterior field is
\begin{equation*}
\label{2.8}
\hat{\rho}_\mathcal{D}(x^0)=\hat{\rho}_0 - \it i g \int d^4y
\theta(x^0-y^0) [\hat{J}_\mu(y),\hat{\rho}_0]a^\mu(y).
\end{equation*}
% corrections
Averaging equations (\ref{1.2}-\ref{1.5}) by means of $\hat{\rho}_\mathcal{D}$ one 
find that the mean values
\[
A_\mu(x) = <\hat{A}_\mu(x)> \equiv A^a_\mu(x)T^a
\]
\noindent and
\[
F_{\mu\nu}(x) = \partial_\mu A_\nu(x) - \partial_\nu A_\mu(x) + 
i g[A_\mu(x), A_\nu(x)] \equiv F^a_{\mu\nu}(x)T^a
\]

\noindent obey the equations
\begin{eqnarray}
\label{2.9}
\mathcal{D}^\mu(A) F_{\mu\nu} & = & g J_\nu - g S_\nu, \\
\label{2.10}
\mathcal{D}^\mu(A) *F_{\mu\nu} & = & g R_\nu,
\end{eqnarray}

\noindent where

\begin{eqnarray}
%\label{2.11}
J_\nu & = & <\hat{J}_\nu> \equiv J^a_\nu T^a, \\
\label{2.12}
S_\nu & = & i \{ \partial^\mu (<[\hat{A}_\mu,\hat{A}_\nu]> - [A_\mu,A_\nu]) + \\
 & & + <[\hat{A}^\mu,\hat{F}_{\mu\nu}]> - [A^\mu,F_{\mu\nu}] \} 
\equiv  S^a_\nu T^a,  \nonumber \\
\label{2.13}
R_\nu & = & \frac{1}{2}\epsilon_{\nu\mu\sigma\tau} i \{ \partial^\mu 
(<[\hat{A}^\sigma,\hat{A}^\tau]> - [A^\sigma,A^\tau]) + \\
& & + <[\hat{A}^\mu,\hat{F}^{\sigma\tau}]> 
- [A^\mu,F^{\sigma\tau}] \} \equiv  R^a_\nu T^a \nonumber
\end{eqnarray}
\noindent (the same notation for operator commutator and Lie algebra commutator
 should not be confusing). The dual of a tensor $F_{\mu\nu}$ is 
$*F_{\mu\nu} \equiv \frac{1}{2}\epsilon_{\mu\nu\sigma\tau}F^{\sigma\tau}$ 
($\epsilon_{0123} = -\epsilon^{0123} = -1$) and $\mathcal{D}^\mu(A)F_{\mu\nu} 
= \partial^\mu F_{\mu\nu} + ig[A^\mu,F_{\mu\nu}]$.
\subsection{Retarded Green's functions and tensors of conductivity}
For the mean value of an operator $A$ one obtains in the first approximation
\begin{eqnarray}
\label{2.14}
<\hat{A}(x^0,\vec{x})> \equiv tr(\hat{A}(x^0,\vec{x}) \rho_{\mathcal{D}}(x^0))
=  \nonumber \\ 
= <\hat{A}(x^0,\vec{x})>_0 - ig \int_{-\infty}^{+\infty} dt 
<[\hat{A}(x^0,\vec{x}),\hat{V}_{\mathcal{D}}(x^0)]>_0\theta(x^0-t)
\end{eqnarray}

\noindent Using the form of $\hat{V}_\mathcal{D}$ from equation (\ref{2.5}) 
and the notation
\[
G_{A,J_\mu}(x,y) = -i\theta(x^0-y^0)<[\hat{A}(x),\hat{J}_\mu(y)]>_0
\]

\noindent for the retarded Green's function, one rewrites equation (\ref{2.14})
 as

\begin{equation}
\label{2.15}
<\hat{A}(x)>=<\hat{A}(x)>_0+g\int d^4y G_{A,J_\mu}(x,y) a^\mu(y).
\end{equation}

For two operators $\hat{A}(x)$ and $\hat{B}(x)$ which are covariant under space 
translations the corresponding Green's function $G_{AB}(x,y) = 
-i \theta(x-y)<[\hat{A}(x),\hat{B}(y)]>_0$ depends only on $x-y$ and in this
case we shall simply write $G_{AB}(x,y) \equiv G_{AB}(x-y)$. Let $\hat{A}$
and $\hat{B}$ transform under the time reflection antiunitary operator $U$
according to

\[
U\hat{A}(x)U^{-1} = \epsilon_A \hat{A}^*(i_tx), 
U\hat{B}(x)U^{-1} = \epsilon_B \hat{B}^*(i_tx)
\]

\noindent where $^*$ denotes hermitian conjugation, $i_tx = (-x^0,\vec{x})$, 
and $|\epsilon_A|=|\epsilon_B|=1$. If, in addition, $\hat{A}$ and $\hat{B}$
are hermitian, the invariance under time reflection and space translations
implies the property

\begin{equation}
\label{2.16}
G_{AB}(k_0,\vec{k}) = \epsilon_A\epsilon_B G_{AB}(k_0,-\vec{k})
\end{equation}

\noindent of the Fourier transform of $G_{AB}$

\[
G_{AB} = \int e^{ik_ox_0-i\vec{k}.\vec{x}} G_{AB}(x_0,\vec{x}) dx_od^3x
\]

\noindent (denoted by the same symbol). Substituting in equation (\ref{2.16})
the components $\hat{J}_\mu^a(x)$ of the fermionic current from (\ref{2.6}),
(\ref{1.7}) we obtain

\begin{equation}
\label{2.17}
J_\mu^a(x) = <\hat{J}_\mu^a(x)> = g \int d^4y G_{\mu\nu}^{ab}(x-y)a^{b,\nu}(y),
\end{equation}

\noindent $a=1, ..., r$, where

\begin{equation*}
\label{2.18}
G_{\mu\nu}^{ab}(x-y) = -i \theta(x^0-y^0)<[\hat{J}_\mu^a(x), 
\hat{J}_\nu^b(x)]>_0
\end{equation*}

\noindent and we have used the fact that for a neutral medium in equilibrium 
the mean values $<\hat{J_\mu^a(x)}>_0 = 0$.

The mean values which enter $S_\nu$ and $R_\nu$ may be expressed by the
corresponding Green's functions, namely

\begin{eqnarray}
\label{2.19}
i\partial^\mu<[\hat{A}_\mu,\hat{A}_\nu]^a> + 
i<\hat{A}^\mu,\hat{F}_{\mu\nu}]^a>  =  g \int d^4y K_{\nu\lambda}^{ab}(x-y) 
a^{b,\lambda}(y), &&\\
\label{2.20}
\frac{1}{2}\epsilon_{\nu\mu\sigma\tau}i\{ \partial^\mu 
<[\hat{A}^\sigma,\hat{A}^\tau]^a> + <\hat{A}^\mu,\hat{F}_{\sigma\tau}]^a>\}  
=&& \nonumber \\ 
= g\int d^4y L_{\nu\lambda}^{ab}(x-y)a^{b,\lambda}(y), &&
\end{eqnarray}

\noindent with
\begin{eqnarray*}
%\label{2.21}
K_{\nu\lambda}^{ab}(x-y) = -i\theta(x_0-y_0)i 
\biggl\{\partial^\mu<\Bigl[[\hat{A}_\mu(x),\hat{A}_\nu(x)]^a,
\hat{J}_\lambda^b(y)\Bigr]>_0 +\nonumber \\
+<i\Bigl[[\hat{A}^\mu(x),\hat{F}_{\mu\nu}(x)]^a,\hat{J}_\lambda^b(y)
\Bigr]>_0\biggr\}, \\
%\label{2.22}
L_{\nu\lambda}^{ab}(x-y) = -i\theta(x_0-y_0)\frac{i}{2} 
\epsilon_{\nu\mu\sigma\tau} \biggl\{
\partial^\mu<\Bigl[[\hat{A}^\sigma(x),\hat{A}^\tau(x)]^a,\hat{J}_\lambda^b(y)
\Bigr]>_0 +
\nonumber\\
+<\Bigl[[\hat{A}^\mu(x),\hat{F}^{\sigma\tau}(x)]^a,\hat{J}_\lambda^b(y)
\Bigr]>_0\biggr\}.
\end{eqnarray*}

The mean values (\ref{2.19}) and (\ref{2.20}) evaluated by means of 
$\hat{\rho}_0$ vanish because Lorentz covariance and invariance under space 
translations imply that there is no non-zero 4-(pseudo) vector with values in 
the Lie algebra $\mathcal{G}$. The components $G_{\mu\nu}^{ab}$ in equation
(\ref{2.17}) will be said to form the exterior conductivity tensor, by 
analogy with classical electrodynamics. Since the current $S_\nu$ and $R_\nu$
are due to the non-Abelian character of the gauge group $G$, the components 
$K_{\nu\lambda}^{ab}$ and $L_{\nu\lambda}^{ab}$ will be said to form the
non-Abelian conductivity tensor and pseudotensor, respectively.

\section{Tensors of generalized susceptibilities}
Proceeding in analogy with classical electrodynamics \cite{22} we introduce
the analog of polarization tensor

\[
\mathscr{P}_{\mu\nu}(x) =
-\mathscr{P}_{\nu\mu}(x)=\mathscr{P}^a_{\mu\nu}T^a,
\]

\noindent through a gauge covariant relation between $\mathscr{P}_{\mu\nu}$
and the averaged current $J_\nu$,

\begin{equation}
\label{3.1}
J_{\nu}(x) = \mathcal{D}^\mu\mathscr{P}_{\mu\nu}(x) =
\partial^\mu\mathscr{P}_{\mu\nu}(x) -
\it i g[A^\mu(x),\mathscr{P}_{\mu\nu}(x)],
\end{equation}

\noindent the covariant differentiation being with respect to 
$A_\mu = <\hat{A}_\mu>$. Equation (\ref{3.1}) is the substitute of 
$J_\nu = \partial^\mu P_{\mu\nu}$ in the case of electrodynamics. We define 
also the corresponding generalizations of the electric and magnetic 
polarizations, respectively, as $\mathcal{G}$-valued 3-vectors $\vec{P}$ 
and $\vec{M}$,

\begin{eqnarray*}
P^{a,n} &=& \mathscr{P}^{a,0n}, \\
M^{a,n} &=& -\frac{1}{2}\epsilon^{nlj}\mathscr{P}^{a,lj}.
\end{eqnarray*}

\noindent ($a=1, ..., r$). Written in components, the defining equation 
(\ref{3.1}) reads

\begin{eqnarray}
J^{a,n} &=& \partial_0P^{a,n} + \epsilon^{nmj}(\partial_m M^{a,j} + \it  g
f^{abc}A^{b,m}M^{c,j})\equiv \nonumber \\
\label{3.2}
& & \equiv \partial_0P^{a,n} +
(\vec{\mathcal{D}}(A)\times\vec{M})^{a,n}, \\
\label{3.3} 
J^{a,0} &=& - \vec{\nabla}\cdot\vec{P}^a - \it  g
f^{abc}(A^{b,n}P^{c,n}) \equiv -
(\vec{\mathcal{D}}(A)\cdot\vec{P})^a,
\end{eqnarray}

\begin{eqnarray*}
\vec{\mathcal{D}}(A)\cdot\vec{P}^a = \partial_nP^{a,n} + f^{abc}A^{b,n}
P^{c,n}, \\
\vec{\mathcal{D}}(A)\times\vec{M})^{a,n} = \epsilon^{nkl}
[\partial_kM^{a,l}+f^{abc}A^{b,k}M^{c,l}]
\end{eqnarray*}

\noindent ($\vec{P}^a=(P^{a,n}), \vec{M}^a=(M^{a,n})$).
Applying again the analogy with
the case of electrodynamics, we fix the arbitrariness in $\mathscr{P}_{\mu\nu}$ 
assuming a linear relation between (the Fourier transformations of) 
$\vec{P}^a$ and $\vec{M}^a$, on the one hand, and the averaged "electric" and 
 "magnetic" fields, $\vec{E}^a$ and $\vec{B}^a$, on the other:

\begin{eqnarray}
\label{3.4}
\vec{P}^a(k_0,\vec{k}) &=&
\kappa^{ab}(k_0,\vec{k})\vec{E}^b(k_0,\vec{k}), \\
\label{3.5}
\vec{M}^a(k_0,\vec{k}) &=& \psi^{ab}((k_0,\vec{k})\vec{B}^b(k_0,\vec{k})
\end{eqnarray}

\noindent ($E^{a,n}=-F^{a,0n}$, $B^{a,j}=-\frac{1}{2}\epsilon^{jkn}F^{a,kn}$).
Equations (\ref{3.4}) and (\ref{3.5}) are the analogs of the
constitutive relations of a medium, where the "color" tensor
$\kappa=(\kappa^{ab})$  appear as generalization of the {\it electric susceptibility}. 
The procedure of introducing the polarization tensor may be applied to the 
currents $S_{\nu}$ and $R_{\nu}$ which appear due to the non-Abelian 
character of the group $G$. We define the {\it non-Abelian polarization tensor}

\begin{equation}
\label{3.6}
Q_{\mu\nu}(x) = -Q_{\nu\mu}(x) = Q_{\mu\nu}^aT^a
\end{equation}

\noindent through the gauge covariant relation with the current $S_\nu$:

\[
-S_\nu(x)=\mathcal{D}^\mu(A)Q_{\mu\nu}(x).
\]

Related with $Q_{\mu\nu}$ are the $\mathcal{G}$-valued 3-vectors $\vec{U}$ and
 $\vec{V}$, 

\[
U^{a,n} \equiv Q^{a,0n}, V^{a,n} \equiv \frac{1}{2}\epsilon^{njl}Q^{a,jl}.
\]

\noindent The vectors $\vec{U}^a$ and $\vec{V}^a$ have no analogs in 
electrodynamics and
will be called vectors of {\it non-Abelian electric} and {\it magnetic 
polarization}, respectively. We have (cf. (\ref{3.2}), (\ref{3.3}))

\begin{eqnarray*}
%\label{3.8}
S_0^a &=& (\vec{\mathcal{D}}(A)\cdot\vec{U})^a, \\
%\label{3.9}
\vec{S}^a &=& -\partial_0\vec{U}^a - (\vec{\mathcal{D}}(A)\times\vec{V})^a, 
\end{eqnarray*}

Due to the lack of analogy with the electrodynamical case we may assume either 
a linear relation between $\vec{U}^a$, $\vec{M}^a$, and the fields $\vec{E}^a$, 
$\vec{B}^a$, of the type

\begin{equation}
\label{3.10}
\vec{U}^a(k) = t^{ab}\vec{E}^b(k), \vec{V}^a(k) = u^{ab}(k)\vec{B}^b(k).
\end{equation}

\noindent or a non-linear (e.g. quadratic) relation among them. Here we 
consider the assumption of linear relation and call $t(k)$ tensor of {\it 
non-Abelian electric susceptibility}. The structure of the current $R_\nu(x)$ yields 
the relation

\begin{equation}
\label{3.11}
R_\nu(x) = \mathcal{D}^\mu(A) *Q_{\mu\nu}(x)
\end{equation}

\noindent with the dual of the tensor $Q_{\mu\nu}$, so 

\begin{equation*}
\label{3.12}
R_0^a = - (\vec{\mathcal{D}}(A)\cdot\vec{V})^a, 
\vec{R}^a = \partial_0\vec{V}^a + (\vec{\mathcal{D}}(A)\times\vec{V})^a.
\end{equation*}

\section{In-medium Yang-Mills equations}

Inserting the expressions for the currents $J_\nu$, $R_\nu$ and $S_\nu$ in terms
 of the susceptibility tensors allows one to recast the averaged system of
equations (\ref{2.9}), (\ref{2.10}) in the form

\begin{eqnarray}
i k^m(\delta^{ab}+g\kappa^{ab}(k)+gt^{ab}(k))E^{b,m}(k)+ && \nonumber \\
+g f^{abc}\int A^{b,m}(k-q)(\delta^{cd}+g\kappa^{cd}(q)+gt^{ab}(q))
E^{d,m}(q)\frac{d^4q}{(2\pi)^4} =&& \nonumber \\
\label{4.1}
= g j_o^a(k), &&\\
i k_0(\delta^{ab}+g\kappa^{ab}(k)+gt^{ab}(k))E^{b,m}(k) + & & \nonumber \\
+ \epsilon^{nml}ik^m(\delta^{ab}-g\psi^{ab}(k)+gu^{ab}(k))B^{b,l}(k) + & & 
\nonumber \\
+\epsilon^{nml} g f^{abc}\int A^{b,m}(k-q)(\delta^{cd}
-g\psi^{ab}(q)+gu^{ab}(q))B^{d,l}(q)\frac{d^4q}{(2\pi)^4} =&& \nonumber \\
\label{4.2}
 = g j^{a,n}(k),\\
\label{4.3}
ik^m(\delta^{ab}-gu^{ab}(k))B^{b,m}(k)+\nonumber\\
gf^{abc}\int A^{b,m}(k-q)
(\delta^{cd}-gu^{cd}(q))B^{m,d}(q)\frac{d^4q}{(2\pi)^4}=0,\\
\label{4.4}
ik_0(\delta^{ab}-gu^{ab}(k))B^{b,n}(k)+\epsilon^{nlm}ik^l(\delta^{ab}-
gt^{ab}(k))E^{b,m}(k)+\nonumber\\
+gf^{abc}\int A^{b,l}(k-q)(\delta^{cd}-gt^{cd}(q))E^{m,d}(q)
\frac{d^4q}{(2\pi)^4}=0.
\end{eqnarray}
Let us define the {\it modified "color" electric field} $\vec{E}'^a$
\begin{equation*}
\label{4.5}
\vec{E}'^a(k)\equiv \vec{E}^a(k)-g\vec{U}^a(k)
\end{equation*}

\noindent and the {\it modified "color" magnetic field} $\vec{B}'^a$

\begin{equation*}
\label{4.6}
\vec{B}'^a(k)\equiv \vec{B}^a(k)-g\vec{V}^a(k).
\end{equation*}

We define also the {\it generalized electric induction}
\begin{equation*}
\label{4.7}
\vec{D}^a(k)\equiv \vec{E}^a(k)+g\vec{P}^a(k)+g\vec{U}^a(k)
\end{equation*}

\noindent and the {\it generalized magnetic induction}
\begin{equation}
\vec{H}^a(k)\equiv \vec{B}^a(k)-g\vec{M}^a(k)+g\vec{V}^a(k).
\end{equation}

\noindent From (\ref{3.10}) one obtains

\begin{eqnarray}
\label{4.8}
\vec{E}'^a(k) &=& (\delta^{ab}-gt^{ab}(k))\vec{E}^b(k) \nonumber \\
\\
\vec{B}'^a(k) &=& (\delta^{ab}-gu^{ab}(k))\vec{B}^b(k) \nonumber
\end{eqnarray}

\noindent and from (\ref{3.4}),(\ref{3.5}),(\ref{3.11}) we get

\begin{eqnarray}
\vec{D}^a(k)&=&(\delta^{ab}+g\kappa^{ab}(k)+gt^{ab}(k))\vec{E}^b(k),\nonumber \\
\vec{H}^a(k)&=&(\delta^{ab}-g\psi^{ab}(k)+gu^{ab}(k))\vec{B}^b(k). 
\nonumber
\end{eqnarray}

\noindent We write these relations in the form

\begin{eqnarray}
\label{4.9}
\vec{D}^a(k)&=&\epsilon^{ab}(k)\vec{E}^b(k) \nonumber\\
\\
\vec{B}^a(k)&=&\mu^{ab}(k)\vec{H}^b(k), \nonumber
\end{eqnarray}

\noindent where

\[
\epsilon^{ab}(k)\equiv \delta^{ab}+g\kappa^{ab}(k)+gt^{ab}(k)
\]

\noindent will be called tensor of {\it generalized electric permittivity}, and

\[
\mu^{ab}(k)\equiv((1-g\psi+gu)^{-1})^{ab}
\]

\noindent - tensor of {\it generalized magnetic permeability}. If we keep the
analogy with the electromagnetic properties of the vacuum, we have to prescribe 
the property of generalized permeability $\mu_0$ and permittivity $\epsilon_0$ 
to the empty space with $\mu_0=const$, $\epsilon_0=const$. Then the definitions 
of $\mu$ and $\epsilon$ will read

\[
\mu^{ab}/\mu_0 \equiv \delta^{ab}+g\chi^{ab}, \mbox{ and }
\epsilon^{ab}/\epsilon_0 \equiv \delta^{ab}+g\kappa^{ab}.
\]

Now the system (\ref{4.1})-(\ref{4.4}) written as a system of differential 
equations reads
\begin{eqnarray}
\label{4.10}
(\vec{\mathcal{D}}(A)\cdot\vec{D})^a = g j_0^a, \nonumber\\
(\vec{\mathcal{D}}(A)\times\vec{H})^a - \partial_0\vec{D}^a = g\vec{j}^a
\nonumber\\
\\
(\vec{\mathcal{D}}(A)\cdot\vec{B}')^a = 0, \nonumber \\ 
(\vec{\mathcal{D}}(A)\times\vec{E}')^a - \partial_0\vec{B}'^a = 0. \nonumber
\end{eqnarray}

The system (\ref{4.10}) together with equations (\ref{4.8}) and (\ref{4.9})
where the tensors $t^{ab}$, $u^{ab}$, $\epsilon^{ab}$ and $\mu^{ab}$ are given 
functions of $(x^0,\vec{x})$ will be called system of {\it in-medium 
Yang-Mills equations}.

Our next step is to outline how relations between the susceptibility tensors
and the Green's functions can be derived in principle. The Fourier transform 
of the system (\ref{1.1}) written in terms of the fields 
$e^{c,n} = -f^{c,0n}$ and $b^{c,n}=-\frac{1}{2}\epsilon^{njm}f^{c,jm}$,
 ($c=1, ...,r; n=1,2,3$) and restricted to terms linear in the external field
gives for the external current

\begin{eqnarray}
\label{5.1}
gj_0^a(k) &=& i\vec{k}\cdot\vec{e}^a(k) \nonumber \\
\\
g\vec{j}^a(k)&=&i\frac{k^2}{k_0}\vec{e}^a(k)+i\frac{\vec{k}\cdot\vec{e}^a(k)}
{k_0}\vec{k} \nonumber
\end{eqnarray}

\noindent ($k=(k_0,\vec{k})$, $k^2=k_0^2-\vec{k}^2$). The substitution of
(\ref{5.1}) into the r.h.s of the system of equations (\ref{4.1})-
(\ref{4.4}) gives a non-linear system for the fields $\vec{E}^a(k)$ and 
$\vec{B}^a(k)$ (recall that $\vec{E}^a(k)=ik_0\vec{A}^a(k)$ in the gauge
$\hat{A}_0=0$). In the linear approximation with respect to the external field 
we have to keep only the linear terms in the expressions for $\vec{E}^a$ and 
$\vec{B}^a$,

\begin{equation}
\label{5.2}
\vec{E}^a(k)=\phi^{ab}\vec{e}^b(k), \vec{B}^a(k)=\theta^{ab}\vec{e}^b(k).
\end{equation}

Here $\phi^{ab}$ and $\theta^{ab}$ depend on $k$ as well as on the tensors 
$\kappa$, $\psi$, $t$ and $u$. The form (\ref{5.2}) implies that 
$[A_\mu,A_\nu]$ and $[A_\mu,F_{\sigma\tau}]$ in equations (\ref{2.12}), 
(\ref{2.13}) are quadratic in the external field and have to be dropped out 
thus simplifying the currents $S_\nu$ and $R_\nu$ to

\begin{eqnarray}
\label{5.3}
S_\nu^a(x) &=& i\Bigl\{
\partial^\mu<[\hat{A}_\mu(x),\hat{A}_\nu(x)]^a>+
<[\hat{A}^\mu(x),\hat{F}_{\mu\nu}(x)]^a>\Bigr\} \nonumber \\
&=& g \int K_{\nu\lambda}^{ab}(x-y)a^{b,\lambda}(y)d^4y \\
\mbox{and}\nonumber\\
\label{5.4}
R_\nu^a(x) &=& \frac{i}{2}\epsilon_{\nu\mu\sigma\tau}\Bigl\{
\partial^\mu<[\hat{A}^\sigma(x),\hat{A}^\tau(x)]^a> + 
<[\hat{A}^\mu(x),\hat{F}^{\sigma\tau}(x)]^a>\Bigr\} \nonumber \\
&=&g\int L_{\nu\lambda}^{ab}(x-y)a^{b,\lambda}(y)d^4y.
\end{eqnarray}

Adding the formula for $J_\nu(x)$,
\begin{equation}
\label{5.5}
J_\nu^a(x)=<\hat{I}_\nu^a(x)>=
g\int G_{\nu\lambda}^{ab}(x-y)a^{b,\lambda}(y)d^4y
\end{equation}

\noindent one may express the three currents (\ref{5.3}), (\ref{5.4}), 
(\ref{5.5}), further in terms of the external field since in any fixed gauge 
for $a^{b,\lambda}$ the equations

\begin{eqnarray}
\label{5.6}
\vec{e}^c(k)&=&ik_0\vec{a}^c(k)-i\vec{k}a^{0,c}(k) \nonumber \\
\\
\vec{b}^c(k)&=&\frac{1}{k_0}\vec{k}\times\vec{e}^c(k)=
i\vec{k}\times\vec{a}^c(k), \nonumber 
\end{eqnarray}

\noindent determine $a^{c,\lambda}$ in terms of $\vec{e}^c$.

On the other hand, the currents $J_\nu$ and $S_\nu$ are used to introduce the 
generalized susceptibility tensors via (\ref{3.1}) and (\ref{3.6}). 
The assumption that these tensors are linearly connected with $\vec{E}^a$ and 
$\vec{B}^a$ (equations (\ref{3.4}), (\ref{3.5}) and (\ref{3.10})) allows one 
to express them linearly by $\vec{e}^b$ using (\ref{5.2}). The obtained two 
expressions for each of the currents $J_\nu$ and $S_\nu$ allow 
one to derive algebraic relations among the tensors $\kappa$, $\psi$, $t$ and 
$u$, and the Green's functions $G_{\nu\lambda}^{ab}$, $K_{\nu\lambda}^{ab}$.

Let us note that if the tensors $t^{ab}$ and $u^{ab}$ depend in a non-linear
(e.g. quadratic) way on the fields $\vec{E}^a$ and $\vec{B}^a$, then they
have no contribution in the averaged system (\ref{4.1}) - (\ref{4.4}) in the
approximation linear in the external field. Then the modified fields 
$\vec{E}'^a$ and $\vec{B}'^a$ reduce to $\vec{E}^a$ and $\vec{B}^a$, 
respectively, and the in-medium Yang-Mills equations become

\begin{eqnarray}
\label{5.7}
(\vec{\mathcal{D}}(A)\cdot\vec{D})^a = gj_0^a, \nonumber \\
(\vec{\mathcal{D}}(A)\times\vec{H})^a -\partial_0\vec{D}^a = g\vec{j}^a, 
\nonumber \\
\\
(\vec{\mathcal{D}}(A)\cdot\vec{B})^a = 0, \nonumber \\
(\vec{\mathcal{D}}(A)\times\vec{E})^a -\partial_0\vec{B}^a = 0. \nonumber 
\end{eqnarray}

\section{Approximation up to second order in g}
Considered as perturbative series in $g$ both Green's functions 
$K_{\nu\lambda}^{ab}$ and $L_{\nu\lambda}^{ab}$ have their first non-vanishing
term of first order in $g$. Indeed, let us take a perturbative expansion of
$\hat{A}_\mu(x)$ and $\hat{F}_{\mu\nu}(x)$ in powers of $g$ beginning with the 
free field operators $\hat{A}_{(0)\mu}$ and $\hat{F}_{(0)\mu\nu}$, 
respectively,

\begin{eqnarray*}
\hat{A}_\mu(x) &=& \hat{A}_{(0)\mu}(x)+\sum_{n\ge 1}g^n\hat{A}_{(n)\mu}(x)\\
\hat{F}_{\mu\nu}(x) &=& \hat{F}_{(0)\mu\nu}(x)+
\sum_{n\ge 1}g^n\hat{F}_{(n)\mu\nu}(x)
\end{eqnarray*}

\noindent ($\hat{A}_{(n)\mu}(x)$ being the $n-th$ approximation). Then the 
observation that $\hat{A}_{(0)\mu}(x)$ and $\hat{F}_{(0)\mu\nu}(x)$ commute
with the current $\hat{I}_\lambda$ expressed in terms of free spinor operators
yields the conclusion that both $K_{\nu\lambda}^{ab}$ and $L_{\nu\lambda}^{ab}$
are of order $g$. According to (\ref{5.3}) and (\ref{5.4}) the terms
with $gS_\nu$ and $gR_\nu$ in the r.h.s of the system of equations (\ref{2.9}), 
(\ref{2.10}), are of third order in $g$ while the term with $gJ_\nu$ is of 
second order (like the case of electrodynamics).

In this Section we consider the systems (\ref{2.9}), (\ref{2.10}), up to second
order in $g$ (in the approximation linear w.r.t. the external field). In this 
approximation the currents $gS_\nu$ and $gR_\nu$ drop out, the covariant 
derivatives in the l.h.s of (\ref{2.9}), (\ref{2.10}) reduce to simple partial
derivatives because $[A^\mu,F_{\sigma\tau}]$ are quadratic in the external
field due to equation (\ref{5.2}), and we obtain the simplified form

\begin{eqnarray*}
\label{6.1}
\partial^\mu\tilde{F}_{\mu\nu}^a = gJ_\nu^a + gj_\nu^a, \nonumber \\
\partial^\mu *\tilde{F}_{\mu\nu}^a = 0,\\
\tilde{F}_{\mu\nu}^a = \partial_\mu A_\nu^a - \partial_\nu A_\mu^a. 
\nonumber 
\end{eqnarray*}

Local commutativity of the currents $J_\nu^a(x)$ gives that

\[
\partial^\mu G_{\mu\nu}^{ab}(x)=0=\partial^\nu G_{\mu\nu}^{ab}(x)
\]

\noindent because $\partial^\mu J_\mu(x)=0$ in our approximation. Using the
Fourier transform of the last equality and the system (\ref{5.6}) (in the gauge
$\partial_\lambda a^\lambda = 0$) we recast equation (\ref{5.5}).

\[
J_\nu^a(k) = g G_{\nu\lambda}^{ab}(k)a^{\lambda,b}(k)
\]

\noindent in the form ($k=(k_0,\vec{k})$)

\begin{eqnarray*}
\label{6.2}
J_0^a(k_0,\vec{k}) &=& ig\frac{k^m}{k_0^2}G_{ml}^{ab}(k_0,\vec{k})
e^{b,l}(k_0,\vec{k}), \nonumber \\
\\
J_n^a(k_0,\vec{k}) &=& -ig\frac{1}{k_0}G_{nl}^{ab}(k_0,\vec{k})
e^{b,l}(k_0,\vec{k}). \nonumber
\end{eqnarray*}

For a homogeneous and isotropic medium $G_{mn}^{ab}(k_0,\vec{k})$ may be 
written as a sum of tensors

\begin{equation*}
\label{6.3}
G_{mn}^{ab}(k_0,\vec{k}) = -ik_0\Bigl[
\tilde{\sigma}_l^{ab}(k)\frac{k_mk_n}{\vec{k}^2}
+\tilde{\sigma}_t^{ab}(k)(\delta_{nm}-\frac{k_mk_n}{\vec{k}^2})
\Bigr],
\end{equation*}

\noindent where $\tilde{\sigma}_l^{ab}$ and $\tilde{\sigma}_t^{ab}$ are 
"color" tensors depending on $k=|\vec{k}|$ and arise as analogs of the so 
called exterior longitudinal and transverse electric conductivity coefficients 
\cite{22}. This form of $G_{mn}^{ab}$ generates a representation of $J_0^a$ 
and $J_n^a$,

\begin{eqnarray}
\label{6.4}
J_0^a(k_0,\vec{k}) &=& \frac{g}{k_0}\tilde{\sigma}_l^{ab}
(\vec{k}\cdot\vec{e}^b) \nonumber \\
\\
J_n^a(k_0,\vec{k}) &=& g \tilde{\sigma}_l^{ab}(k)
\frac{\vec{k}\cdot\vec{e}^b}{\vec{k}^2}k_n + g \tilde{\sigma}_t^{ab}(k)
\frac{((\vec{k}\times\vec{e}^b)\times\vec{k})_n}{\vec{k}^2}, \nonumber 
\end{eqnarray}

\noindent the second equation being a decomposition of $J_n^a$ into
longitudinal and transverse parts with respect to $\vec{k}$. It is convenient 
to write the vectors $\vec{P}^a$ and $\vec{M}^a$, equations (\ref{3.4}), 
(\ref{3.5}), in the form

\begin{eqnarray}
\label{6.5}
\vec{P}^a(k_0,\vec{k}) &=& \kappa^{ab}(k_0,\vec{k})\vec{E}^b(k_0,\vec{k}),
\nonumber \\
\\
\vec{M}^a(k_0,\vec{k}) &=& (\mu^{-1})^{ab}(k_0,\vec{k})\chi^{bc}(k_0,\vec{k})
\vec{B}^c(k_0,\vec{k}), \nonumber 
\end{eqnarray}

\noindent (i.e. we write $\psi = \mu^{-1}\cdot\chi$) where $\mu^{ab}=
\delta^{ab}+g\chi^{ab}$. The Fourier transform of the system (\ref{1.5}) 
written in terms of $E^{e,n}=-\tilde{F}^{a,0n}$ and 
$B^{a,n}=-\frac{1}{2}\epsilon^{nml}\tilde{F}^{a,ml}$ (i.e. the system of
equations (\ref{4.1}) - (\ref{4.4}) in the approximation considered) reads

\begin{eqnarray}
\label{6.6}
ik^m(\delta^{ab}+\kappa^{ab}(k_0,\vec{k}))E^{b,m}(k_0,\vec{k}) = 
gj_0^a(k_0,\vec{k}), \nonumber \\
ik_0(\delta^{ab}+\kappa^{ab}(k_0,\vec{k})E^{b,m}(k_0,\vec{k}) + && \nonumber \\
\epsilon^{nml}ik^m(\delta^{ab}-g(\mu^{-1})^{ab}(k_0,\vec{k})
\chi^{bc}(k_0,\vec{k}))B^{b,l}(k_0,\vec{k}) = gj^{a,n}(k_0,\vec{k}), \\
ik^mB^{a,m}(k_0,\vec{k}) = 0, \nonumber \\
ik_0B^{a,n}(k_0,\vec{k})+\epsilon^{nml}ik^lE^{a,m}(k_0,\vec{k}) = 0. 
\nonumber 
\end{eqnarray}

The vectors of generalized electric and magnetic induction are

\begin{eqnarray*}
\label{6.7}
\vec{D}^a &=& \vec{E}^a+g\vec{P}^a=\epsilon^{ab}\vec{E}^b, \\
\label{6.8}
\vec{H}^a &=& \vec{B}^a-g\vec{M}^a=(\mu^{-1})^{ab}\vec{B}^b,
\end{eqnarray*}

\noindent with $\epsilon^{ab}=\delta^{ab}+g\kappa^{ab}$. In terms of these
fields the form of the system (\ref{6.6}) as a system of differential equations reads

\begin{eqnarray}
\label{6.9}
div\vec{D}^a = j_0^a, \nonumber\\
rot\vec{H}^a - \partial_0\vec{D}^a = g\vec{j}^a, \nonumber \\
\\
div\vec{B}^a = 0, \nonumber \\
rot\vec{E}^a-\partial_0\vec{B}^a = 0. \nonumber 
\end{eqnarray}

It follows from the above equations that

\[
\vec{D}_{||}^c(k_0,\vec{k})\equiv\frac{1}{\vec{k}^2}
(\vec{k}\cdot\vec{D}^c(k_0,\vec{k}))\vec{k}=
\epsilon^{cd}(k_0,\vec{k})\vec{E}_{||}^d(k_0,\vec{k})=
\vec{e}_{||}^c(k_0,\vec{k}),\\
\]
\[
[k_0^2\epsilon^{ab}-\vec{k}^2(\mu^{-1})^{ab}]
\vec{E}_\perp^b(k_0,\vec{k}) = (k_0^2-\vec{k}^2)\vec{e}_\perp^a(k_0,\vec{k})
\]

\noindent and

\[
[k_0^2\epsilon^{ab}-\vec{k}^2(\mu^{-1})^{ab}]
\vec{B}_\perp^b(k_0,\vec{k}) = (k_0^2-\vec{k}^2)\vec{h}^a(k_0,\vec{k}),
\]

\noindent with $\vec{B}_{||}^a=0$, where for a vector $\vec{v}^a$ we define

\[
\vec{v}_\perp^a = \frac{1}{\vec{k}^2}(\vec{k}\times\vec{v}^a)\times\vec{k}.
\]

With the help of equations (\ref{6.4}), (\ref{6.5}), and the relations

\begin{eqnarray*}
J_0^a(k_0\vec{k})&=&-i\vec{k}\cdot\vec{P}^a(k_0,\vec{k}), \\
\vec{J}^a(k_0\vec{k})&=&
-ik_0\vec{P}^a(k_0\vec{k})+i\vec{k}\times\vec{M}^a(k_0\vec{k}),
\end{eqnarray*}

\noindent now we derive

\[
k_0\kappa^{ab}(k_0\vec{k})=ig\tilde{\sigma}_t^{ab}(|\vec{k}|)
\epsilon^{cb}(k_0\vec{k}),
\]

\noindent or, in a compact matrix notation

\begin{equation}
\label{6.10}
1-\epsilon^{-1}=ig^2\tilde{\sigma}_t,
\end{equation}

\noindent as well as the relation

\begin{eqnarray}
\label{6.11}
1-\mu^{-1}=g\mu^{-1}\chi = && \nonumber \\
i\frac{k_0}{\vec{k}^2}g^2 \Bigl\{
\tilde{\sigma}_t(1-ig^2k_0\frac{\tilde{\sigma}_t}{k_0^2-\vec{k}^2})^{-1}
-(1-i\frac{g^2}{k_0}\tilde{\sigma}_l)^{-1}\tilde{\sigma}_l
\Bigr\}.
\end{eqnarray}

Equations (\ref{6.10}) and (\ref{6.11}) resemble their counterparts in
electrodynamics and by analogy, we call

\[
\sigma_l = (1-i\frac{g^2}{k_0}\tilde{\sigma}_l)^{-1}\tilde{\sigma}_l,
\]

\noindent

\[
\sigma_t =\tilde{\sigma}_t(1-ig^2k_0
\frac{\tilde{\sigma}_t}{k_0^2-\vec{k}^2})^{-1},
\]

\noindent tensors of intrinsic longitudinal and transverse conductivity, 
respectively.
%% end corrections

\section{Lagrangian and canonical quantization}
In this section we consider the case when the tensors $t^{ab}$ and
$u^{ab}$, equation (\ref{3.10}), may be neglected, so the in-medium Yang-Mills 
fields are described by the system (\ref{5.7}). 

For a medium in rest with respect to a fixed Lorenz frame we define the 
antisymmetric tensors $H^{a, \mu \nu}$  by \begin{eqnarray}
H^{a, 0 \kappa} & = & - D^{a, \kappa}, \nonumber\\
\label{64}
H^{a, m n} & = & - \epsilon^{mnk}H^{a, k}, n,m = 1,2,3.
\end{eqnarray}
The components of the tensors (\ref{64}) read
\begin{equation}
\label{65} H^{a, \mu \nu} = (\mu^{-1})^{ab}F^{b, \mu \nu} +
(\epsilon - \mu^{-1})^{ab}V^\tau(g^{\mu \sigma}V^\nu - g^{\nu
\sigma}V^\mu)F^b_{\sigma \tau},
\end{equation}
\noindent if the medium moves uniformly with velocity $V^\mu$
$(V^2 = 1)$. For vanishing exterior current $j^a_{\mu}$ the
Lagrangian
\begin{equation}
\label{66} \mathscr{L}_0 = -\frac{1}{4}F^a_{\mu \nu}H^{a, \mu \nu}
\end{equation}
\noindent reproduces the first two equations of the system (\ref{6.9}) 
written in a covariant form
\begin{equation*}
\label{67} D_\mu(A){H}^{a, \mu \nu} = 0,
\end{equation*}
\noindent valid for any reference frame. Since in (\ref{65})
and (\ref{66}) the tensor $F^a_{\mu \nu}$ reads
\begin{equation*}
\label{68} F^a_{\mu \nu} = \partial_{\mu} A^a_{\nu}
-\partial_{\nu} A^a_{\mu}- g f^{abc}A^b_{\mu} A^c_{\nu},
\end{equation*}
\noindent it satisfies the covariant form of the last two
equations of the system (\ref{6.9}), i.e.
\begin{equation*}
\label{69}
D_\lambda(A) F^a_{\mu \nu} + D_\mu(A) F^a_{\nu \lambda} + D_\nu(A)
F^a_{\lambda \mu} = 0.
\end{equation*}
In order to perform the canonical quantization, we add a gauge fixing term
$\mathscr{L}_{g f}$ to $\mathscr{L}_0$,
\begin{equation}
\label{70} \mathscr{L}_{g f} = \frac{\xi}{2}(B^aB^a) -
B^a(\partial^\mu A^a_\mu + (n^2-1)^{ab}V^\mu\partial_\mu V^\lambda
A^b_\lambda),
\end{equation}
\noindent where $\xi$ is a real parameter and $B^a$ $(a=1,...,r)$
are real auxiliary scalar fields. The tensor $(n^2)^{ab} =
\epsilon^{ac}\mu^{cb}$ is the generalization of the classical
refraction coefficient. The form (\ref{70}) of $\mathscr{L}_{g
f}$ is taken in analogy with the case of electromagnetic field in
a medium \cite{23}. In the sequel we will consider the case
$\epsilon^{ab}=\epsilon\delta^{ab}$, $\mu^{ab}=\mu\delta^{ab}$,
where $\epsilon$ and $\mu$ are constants. Then
$(n^2)^{ab}=\epsilon\mu\delta^{ab}$ and the notation $\beta=n^2-1$
will be used (here and below we shall write $n^2$ for
$\epsilon\mu$). We split the total Lagrangian
$\mathscr{L}=\mathscr{L}_0+\mathscr{L}_{g f}$ into two parts,
$\mathscr{L}=\mathscr{L}_{free}+\mathscr{L}_{int}$, where
$\mathscr{L}_{int}$ contains all terms containing the gauge
coupling constant $g$. The part $\mathscr{L}_{free}$ equals
\begin{eqnarray}
\mathscr{L}_{free} = -\frac{1}{4}\tilde{F}^a_{\mu
\nu}\tilde{H}^{a, \mu \nu} + \mathscr{L}_{g f},
\end{eqnarray}
\noindent where
\begin{equation}
\tilde{F}^a_{\lambda\mu}=\partial_\lambda A^a_\mu - \partial_\mu
A^a_\lambda
\end{equation}
\noindent and
\begin{equation}
\tilde{H}^{a, \mu \nu} = (\mu^{-1})^{ab}\tilde{F}^{b, \mu \nu} +
(\epsilon - \mu^{-1})^{ab}V^\tau(g^{\mu \sigma}V^\nu - g^{\nu
\sigma}V^\mu)\tilde{F}^b_{\sigma \tau}.
\end{equation}
The equations of motion following from the free field part
$\mathscr{L}_{free}$ are
\begin{eqnarray}
\label{71} \opensquare A^a_\lambda + \beta(V\cdot
\partial)^2A^a_\lambda+(\mu-\xi)\partial_\lambda B^a \equiv
QA^a_\lambda+(\mu-\xi)\partial_\lambda B^a = 0,  & & \\
\label{72}
\opensquare B^a+\beta(V\cdot\partial)^2B^a \equiv QB^a = 0, & & \\
\label{73} \xi B^a = \partial\cdot A^a +
\beta(V\cdot\partial)(V\cdot A^a), & &
\end{eqnarray}
\noindent where $Q=\opensquare+\beta(V\cdot\partial)^2$. We choose
$A^a_\lambda(x)$ as generalized coordinates, the conjugate
momenta being
\[
\pi^{a, \mu}(x) =
\frac{\partial\mathscr{L}_0}{\partial(\partial_0A^a_\mu)}(x),
\]
\noindent or
\begin{eqnarray*}
\label{74}
\pi^{a, 0} &=& -[1 + \beta V^2_0]B^a,\\
\label{75} \pi^{a, k} &=& \beta V_0V_kB^a +
\frac{\beta}{\mu}V_0V_n\tilde{F}^a_{kn} + [(1+2\beta
V^2_0)\delta_{kn}-\beta V_kV_n]\tilde{F}_{0n},
\end{eqnarray*}
\noindent $(k=1,2,3)$.
From the equal time canonical commutation
relations
\begin{eqnarray*}
[A_\lambda(x),A_\mu(y)]_{x_0=y_0}&=&0, \nonumber \\
\label{76}
[\pi^{a,\lambda}(x),\pi^{b,\mu}(y)]_{x_0=y_0}&=&0,\\
\label{77}
[\pi^{a,\lambda}(x),A^b_\mu(y)]_{x_0=y_0}
&=& -i\delta^{ab}\delta^\lambda_\mu\delta(\vec{x}-\vec{y}),
\end{eqnarray*}
\noindent one derives
\begin{eqnarray*}
\label{78} [B^a(x),A^b_\lambda(y)]_{x_0=y_0} &=&
\frac{i\delta^{ab}\delta_{0\lambda}}{(1+\beta
V^2_0)}\delta(\vec{x}-\vec{y}), \\
\label{79} [B^a(x),\partial_0 A^b_n(y)]_{x_0=y_0} &=&
\frac{-i\delta^{ab}}{(1+\beta V^2_0)}
\frac{\partial}{\partial x^n}\delta(\vec{x}-\vec{y}),
\end{eqnarray*}
\noindent as well as
\begin{eqnarray*}
\label{80} [\partial_0A^a_0(x),A^b_0(y)]_{x_0=y_0} &=&
\frac{i\delta^{ab}\delta(\vec{x}-\vec{y})}{(1+\beta V^2_0)^2}
[\xi-\mu\beta^2\frac{V_0^2\vec{V}^2}{(1+\alpha)}],\\
\label{81} [\partial_0 A^a_0(x),A^b_n(y)]_{x_0=y_0} &=&
-\frac{i\mu\beta
V_0V_n\delta^{ab}}{(1+\beta)(1+\beta V^2_0)}\delta(\vec{x}-\vec{y}),\\
\label{82} [\partial_0B^a(x),A^b_0(y)]_{x_0=y_0} &=& 2i\frac{\beta
V_0}{(1+\beta V^2_0)^2}\delta^{ab}(V_n\frac{\partial}{\partial
x^n})\delta(\vec{x}-\vec{y}) ,\\
\label{83} [\partial_0B^a(x),A^b_n(y)]_{x_0=y_0} &=&
\frac{i\delta^{ab}}{(1+\beta V^2_0)}\frac{\partial}{\partial
x^n}\delta(\vec{x}-\vec{y}).
\end{eqnarray*}
Following \cite{23} we introduce the tensor
\[
b^\sigma_\tau=\delta^\sigma_\tau + (n-1)V^\sigma V_\tau,
\]
\noindent satisfying
\[
(b^l)^\sigma_\tau = \delta^\sigma_\tau + (n^l-1)V^\sigma V_\tau
\]
\noindent for every integer number $l$. In terms of $b^\sigma_\tau$, the
gauge fixing term $\mathscr{L}_{gf}$ reads
\[
\mathscr{L}_{gf}=\frac{\xi}{2}B^aB^a - B^a\partial^\mu(b^2)^\tau_\mu
A_\tau
\]
\noindent and the operator Q takes the form
\[
Q = \partial^\mu(b^2)^\tau_\mu\partial_\tau.
\]
\section{Commutation relations at non-equal times}
We define the functions
\begin{equation}
\label{84} F(x)\equiv{i}\int\frac{d^4k}{(2\pi)^3}\exp(-i k
\cdot x)\delta(Q(k))\epsilon(V \cdot k)
\end{equation}
\noindent and
\begin{equation*}
\label{85}
H(x)\equiv{i}\int\frac{d^4k}{(2\pi)^3}\exp(-i k \cdot
x)\delta'(Q(k))\epsilon(V \cdot k),
\end{equation*}
\noindent with the notations
\begin{equation}
Q(k)\equiv k^2+\beta(V \cdot k)^2 = (1+\beta
V^2_0)(k_0-k_+(k))(k_0-k_-(k))
\end{equation}
\noindent and
\begin{equation}
\label{87} k_\pm(\vec{k})\equiv\frac{1}{(1+\beta V^2_0)}\biggl[\beta
V_0(\vec{k}\cdot\vec{V})\pm\sqrt{(1+\beta V^2_0)\vec{k}^2 -
\beta(\vec{k}\cdot\vec{V})^2}\biggr].
\end{equation}
Note that $k_+(\vec{k})$ is negative for reference frames in which 
$\beta \vec{V^2}>1$. However, it always holds that 
$k_{\pm}(\vec{k})=-k_{\mp}(\vec{-k})$. 
  
The functions $F$ and $H$ have the properties $QF=0$, $QH=F$, as well as
\begin{eqnarray}
\partial_0F(0,\vec{x}) =\frac{1}{(1+\beta V^2_0)}\delta(\vec{x}),
\nonumber & & \\
\label{88}
F(0,\vec{x}) = 0, & & \\
H(0,\vec{x})=\partial_0H(0,\vec{x}) = \partial^2_0H(0,\vec{x})=0, \nonumber
& & \\
\label{90}
\partial^3_0H(0,\vec{x}) = \frac{1}{(1+\beta V^2_0)^2}\delta(\vec{x}). & & 
\end{eqnarray}
Equations (\ref{84}) and (\ref{87}) yield
\begin{equation}
F(x)=\frac{i}{(2\pi)^3}\int\frac{d^3k}{(1+\beta V^2_0)}
\frac{1}{|k_+(\vec{k})-k_-(\vec{k})|}(e^{-ik'x}-e^{-ik''x}),
\end{equation}
\noindent where $k'=(k_+(\vec{k}),\vec{k})$,
$k''=(k_-(\vec{k}),\vec{k})$. The current
\begin{equation}
\label{92}
J_\mu(y,x)=F(y-x){\mathscr{D}^{\leftrightarrow}}_{x,\mu}g(x) +
H(y-x){\mathscr{D}^{\leftrightarrow}}_{x,\mu}Qg(x)
\end{equation}
is conserved, $\frac{\partial}{\partial x^\mu}J^\mu(y,x)=0$, for
every solution $g(x)$ of
\[
Q^2 g(x)=0.
\]
\noindent The operator $\mathscr{D}_{x,\mu}$ is defined as
\begin{equation}
\label{93} \mathscr{D}_{x,\mu}\equiv\frac{\partial}{\partial x^\mu} +
\beta V_\mu(V^\lambda\frac{\partial}{\partial x^\lambda})
\end{equation}
\noindent and
\[
A{\mathscr{D}^{\leftrightarrow}}_{x,\mu}B \equiv A(\mathscr{D}_{x,\mu}B)-
(\mathscr{D}_{x,\mu}A)B.
\]
\noindent Therefore the integral
\[
\int_\Sigma d\sigma^\mu(z)J_{\mu}(y,z)
\]
\noindent ($d\sigma^\mu(z)$ is the element on the surface
$\Sigma$) depends only on $y$. For a space-like $\Sigma$ with
$d\sigma^\mu(z)=(1,0,0,0)d^3z$ we define
\begin{equation}
G(y)\equiv\int_{z^0=y^0}d^3z J_0(y,z).
\end{equation}
It follows from equation (\ref{92}) and the properties (\ref{88}) -
(\ref{90}) that
\begin{equation}
G(y)=g(y)=\int_{z^0=y^0}d^3z\biggl[F(y-z)
{\mathscr{D}^{\leftrightarrow}}_{z,0}g(z) +
H(y-z){\mathscr{D}^{\leftrightarrow}}_{z,0}Qg(z)
\biggr].
\end{equation}
From (\ref{71}) and (\ref{72}) one derives
\begin{equation}
\label{94}
Q^2A^a_\lambda(x)=0.
\end{equation}
Therefore we obtain
\begin{equation}
A^a_\lambda(x)=\int_{x^0=z^0}d^3z
[F(x-z){\mathscr{D}^{\leftrightarrow}}_{z,0}A^a_\lambda(z) +
H(x-z)){\mathscr{D}^{\leftrightarrow}}_{z,0}QA^a_\lambda(z)].
\end{equation}
This expression allows one to calculate the commutators among the
fields $A^a_\lambda(x)$ for arbitrary times using the equal time
commutators appearing in the r.h.s of the identify
\begin{eqnarray*}
\label{98} [A^a_\lambda(x),A^b_\mu(y)] &=&
\int_{z_0=y_0}d^3z\biggl\{
F(x-z){\mathscr{D}^{\leftrightarrow}}_{z,0}[A^a_\lambda(z),A^b_\mu(y)]
 \\ \nonumber & &
+ H(x-z){\mathscr{D}^{\leftrightarrow}}_{z,0}(\xi-\mu)
\frac{\partial}{\partial z^\lambda}[B^a(z),A^b_\mu(y)] \biggr\}.
\end{eqnarray*}
The result reads
\begin{equation}
\fl \label{99} [A^a_\lambda(x),A^b_\mu(y)] = 
i\mu\delta^{ab}(g_{\lambda\mu} - \frac{\beta}{1+\beta}V_\lambda
V_\mu)F(x-y) -
i(\mu-\xi)\delta^{ab}\frac{\partial^2H(x-y)}{\partial
x^\lambda\partial x^\mu}.
\end{equation}
Using equation (\ref{73}), i.e.
\[
\xi B^a(x)=\mathscr{D}_{x,\mu}A^{a, \mu}(x),
\]
\noindent with $\mathscr{D}_{x, \mu}$ given by (\ref{93}), one
obtains from equation (\ref{98})
\begin{equation}
\label{100}
[A^a_\lambda(x), B^b(y)]=-i\delta^{ab}\frac{\partial}{\partial x^\lambda}
F(x-y)
\end{equation}
\noindent and
\begin{equation}
\label{101}
[B^a(x),B^b(y)]=0.
\end{equation}
The commutation relations (\ref{99}-\ref{101}) generalize those
derived for the electromagnetic filed in a medium \cite{23} in a
natural way.

One may use the technique of \cite{23} to derive the commutation
relations (\ref{99}) in a much simpler way. Indeed, the change
of variables
\begin{eqnarray*}
\label{102}
x^\lambda &\rightarrow& y^\lambda=(b^{-1})^\lambda_\mu x^\mu, \\
\label{103}
A^a_\lambda(x) &\rightarrow&
\Gamma^a_\lambda(y)=\frac{1}{\sqrt{\mu}}A^a_\sigma(x)b^\sigma_\lambda,\\
B^a(x) &\rightarrow& \Delta^a(y)=\sqrt{\mu}B^a(x) \nonumber
\end{eqnarray*}
and
\begin{equation*}
\label{104}
g \rightarrow g'=\sqrt{\mu}g, \, \xi \rightarrow \xi'=\xi/\mu,
\end{equation*}
brings the Lagrangian
\begin{equation*}
\label{105}
\fl  \mathscr{L} = -\frac{1}{4}F^{a, \mu\nu}(x)\mathscr{H}^a_{\mu\nu}(x) - \xi B^a(x) [\partial \cdot A^a(x) + (n^2-1)(V\cdot\partial)V \cdot A^a(x)]
+\frac{1}{2}B^a(x)B^a(x)
\end{equation*}
to the form
\begin{equation*}
\label{106}
\mathscr{L}=-\frac{1}{4}G^{a, \mu\nu}(y)G^a_{\mu\nu}(y) - \xi'\Delta^a(y)
\frac{\partial}{\partial y^\mu}\Gamma^{a, \mu}(y) + \frac{1}{2}
\Delta^a(y)\Delta^a(y),
\end{equation*}
where
\begin{equation*}
\label{107}
G^a_{\mu\nu}(y)=\frac{\partial\Gamma^a_\nu(y)}{\partial y^\mu} -
\frac{\partial\Gamma^a_\mu(y)}{\partial y^\nu} -
g'f^{abc}\Gamma^b_\mu(y)\Gamma^c_\nu(y).
\end{equation*}
In the approximation where one takes $g' \rightarrow 0$ this
Lagrangian describes a system of $r = dim (G)$ non-interacting
$U(1)$-gauge fields, i.e. a system of $r$ free "electromagnetic"
fields with the corresponding gauge fixing terms containing the
fields $\Delta^a$. After quantization of each type
$\Gamma^a_\mu(x), (a=1,...,r),$ of "electromagnetic" fields one
obtains the commutation relations \cite{23}
\[
[\Gamma^a_\lambda(y), \Gamma^b_\nu(y')] = ig_{\lambda\nu}\delta^{ab}
\mathscr{D}_0(y-y') - i(1-\xi')\frac{\partial^2}{\partial y^\lambda
\partial y^\nu}E(y-y'),
\]
where
\begin{eqnarray*}
\label{109} D_0(y) &=& i\int\frac{d^4k}{(2\pi)^3}e^{-i ky}
\epsilon(k_0)\delta(k^2),\\
E(y) &=& i\int\frac{d^4k}{(2\pi)^3}e^{-i ky}\epsilon(k_0)\delta'(k^2).
\nonumber
\end{eqnarray*}
Going back to the fields $A^a_\lambda$ and the variables
$(x^\mu)$, one recovers equation (\ref{99}) having in mind that 
$D_0(b^{-1} \cdot x) = F(x)$, and $ E(b^{-1}\cdot x)=H(x)$.

\section{Physical subspace of the Fock space}
We introduce the Fourier integral for the fields $A^d_\lambda(x)$ and
$B^a(x)$,
\begin{eqnarray*}
\label{111}
A^d_\lambda(x) &=& \frac{1}{(2\pi)^{3/2}} \int d^4p \theta(p_0)
\biggl[ e^{ipx}{a^+_\lambda}^d(p) + e^{-ipx}{a^-_\lambda}^d(p)\biggr], \\
\label{112}
B^a(x) &=& \frac{1}{(2\pi)^{3/2}} \int d^4p \theta(p_0)
\biggl[ e^{ipx}b^{+\,a}(p) + e^{-ipx}b^{-\,a}(p)\biggr].
\end{eqnarray*}
As a consequence of the equations of motion one obtains
\begin{eqnarray}
%\label{113}
\theta(k^0)[Q(k){a^\pm_\lambda}^d(k) \pm i(\mu-\xi)b^{\pm\,d}(k)] = 0,\\
%\label{114}
\theta(k^0)Q(k)b^{\pm\,d}(k) = 0, \nonumber \\
\label{115}
\pm i\theta(k^0)[k^\lambda {a^\pm_\lambda}^d(k) + (V \cdot k)V^\lambda
{a^\pm_\lambda}^d(k)] = \theta(k^0)\xi b^{\pm\,d}(k). 
\end{eqnarray}
Inserting (\ref{111}) into (\ref{99}) gives the commutation
relations
\begin{equation*}
\label{116} [{a^+_\lambda}^d(k), {a^+_\mu}^c(q)]=0, \,\,
[{a^-_\lambda}^d(k),{a^-_\mu}^c(q)]=0
\end{equation*}
and
\begin{eqnarray*}
\label{117}
[{a^-_\sigma}^d(p), {a^+_\tau}^c(q)] &=& -\mu(g_{\sigma\tau} -
\frac{\beta}{1+\beta}V_\sigma V_\tau)\delta^{dc}\delta(Q(p))\delta(p-q)\\
& & -(\mu - \xi)\delta^{dc}p_\sigma p_\tau \delta'(Q(p))\delta(p-q).\nonumber
\end{eqnarray*}
For the operators $b^{\pm\,d}(p)$ one finds
\begin{equation*}
\label{118}
[b^{+\,d}(p),b^{+\,c}(q)]=0,\,\,[b^{-\,d}(p),b^{-\,c}(q)]=0
\end{equation*}
and
\begin{equation*}
\label{119}
[{a^-_\lambda}^d(p),b^{+\,c}(q)]=-\delta^{dc}p_\lambda\delta(Q(p))\delta(p-q).
\end{equation*}
It is convenient for each four vector $p$ to introduce a local frame
consisting of the ortho-normal vectors $V$, ($V^2=1$), $N(p)$ and
$e^{(r)}(p)$, ($r=1,2$), satisfying
\begin{equation*}
N(p)^2=(e^{(r)}(p))^2=-1,\,\,
\end{equation*}
where
\begin{equation*}
\label{120}
N(p)_\mu\equiv\frac{\tilde{p}_\mu-(V \cdot
\tilde{p})V_\mu}{\sqrt{(V \cdot \tilde{p})^2-\tilde{p}^2}}
\end{equation*}
and $\tilde{p}_\mu\equiv p_\mu+\beta V_\mu(V \cdot p)$. We decompose
the operators ${a^\pm_\lambda}^d(p)$ with respect to the base
vectors $V$, $N(p)$, $e^{(r)}(p)$,
\begin{equation*}
\label{121}
{a^\pm_\lambda}^d(p) = \sum_{r=1}^2 {\alpha^\pm_r}^d(p)e^{(r)}_\lambda(p)
+ {\alpha^\pm_3}^dN_\lambda + {\alpha^\pm_0}^dV_\lambda.
\end{equation*}
The commutation relations for the new operators
${\alpha^\pm_\lambda}^d(p)$ read
\begin{eqnarray}
 \fl \label{122} [{\alpha^-_0}^d(p),{\alpha^+_0}^c(q)] =
-\frac{\mu\delta^{dc}}{1+\beta}\delta(Q(p))\delta(p-q) -
(\mu-\xi)\delta^{dc}(V \cdot p)^2\delta'(Q(p))\delta(p-q),\\
\fl \label{123}
[{\alpha^-_3}^d(p),{\alpha^+_3}^c(q)] =
\mu\delta^{dc}\delta(Q(p))\delta(p-q) - (\mu-\xi)\delta^{dc}[(V\cdot
p)^2-p^2]\delta'(Q(p))\delta(p-q),\\
\fl \label{124}
[{\alpha^-_k}^d(p),{\alpha^+_l}^c(q)] =
\mu\delta^{dc}\delta(Q(p))\delta(p-q),\\
\fl \label{125}
[{\alpha^-_0}^d(p),{\alpha^+_3}^c(q)] = -(\mu-\xi)\delta^{dc}\sqrt{(V\cdot
p)^2-p^2}(V\cdot p)\delta'(Q(p))\delta(p-q),\\
\fl \label{126}
[{\alpha^-_0}^d(p),{\alpha^+_k}^c(q)] = 0,\\
\fl \label{127}
[{\alpha^-_3}^d(p),{\alpha^+_k}^c(q)] = 0,\\
\fl \label{128}
[{\alpha^-_k}^d(p),b^{+\,c}(q)] = 0,\\
\fl \label{129}
[{\alpha^-_3}^d(p),b^{+\,c}(q)] = -i\delta^{dc}\sqrt{(V\cdot
p)^2-p^2}\delta(Q(p))\delta(p-q),\\
\fl \label{130}
[{\alpha^-_0}^d(p),b^{+\,c}(q)] =
-i\delta^{dc}(V\cdot p)\delta(Q(p))\delta(p-q).
\end{eqnarray}
The vacuum vector $|0\rangle$ of the Fock space $\mathscr{H}_F$ is defined
by
\[
{\alpha^-_\lambda}^d(p)|0\rangle=0. 
\]
It follows from the commutation relations
(\ref{122})-(\ref{130}) that the operators ${\alpha^+_0}^d(p)$
and ${\alpha^+_3}^d(p)$ do not generate states with positive
definite norm squared due to the terms with $\delta'(Q(p))$. The
physical subspace of $\mathscr{H}_F$ is defined by
\begin{equation*}
\mathscr{H}_{phys}=\bigg\{ |\Psi\rangle \in \mathscr{H}_F; \,\,
b^{-\,d}(p)|\Psi\rangle=0  \biggr\}.
\end{equation*}
According to equation ({\ref{115}}) the condition
$b^{-\,d}(p)|\Psi\rangle=0$ is equivalent to
\[
\tilde{p}{a^-_\lambda}^d(p)|\Psi\rangle=0,
\]
implying
\begin{equation*}
L^{-\,d}(p)|\Psi\rangle=0,
\end{equation*}
where
\begin{equation*}
L^{\pm\,d}(p)=\sqrt{(V\cdot p)^2-p^2}{\alpha^\pm_0}^d(p)-(V\cdot
p){\alpha^\pm_3}^d(p)
\end{equation*}
for any $|\Psi\rangle \in \mathscr{H}_{phys}$. The states with
non-negative norm squared are generated by ${\alpha^+_{1,2}}^d(p)$ and
$L^{+\,d}(p)$ which are the operators commuting with
$b^{-\,c}(q)$. However, the subspace $\mathscr{H}_0$ of
$\mathscr{H}_{phys}$ generated by $L^{+\,d}(p)$ consists of
zero norm vectors. Therefore, the space of vectors with
positive norm squared is defined by the quotient
\begin{equation*}
\mathscr{H}'=\mathscr{H}_{phys}/\mathscr{H}_0.
\end{equation*}
The energy-momentum tensor $T_{\lambda\mu}$ derived from the
Lagrangian $\mathscr{L}=\mathscr{L}_0+\mathscr{L}_{gf}$ is
\begin{equation*}
T_{\lambda\mu}=-\mathscr{H}^a_{\lambda\tau}\partial_\mu
A^{a,\tau} + (b^{-2}\cdot A^a)_\lambda\partial_\mu
B^a-g_{\lambda\mu}\mathscr{L},
\end{equation*}
$(b^{-2})^\sigma_\lambda=(\delta^\sigma_\lambda+\beta V_\lambda
V^\sigma)$, and may be written as
\[T_{\lambda\mu}=
-\mathscr{H}^a_{\lambda\tau}F^{a,\tau}_\mu
+ \frac{1}{4}g_{\lambda\mu}(F^a_{\alpha\beta}H^{a,\beta\alpha})
\]
up to a total divergence $\sim
\partial^\tau(\mathscr{H}^a_{\lambda\tau}A^{a}_\mu)$. For the matrix
elements of the corresponding four-momentum operator one finds
\begin{equation*}
\langle\Phi|P_\lambda|\Phi'\rangle = \int d^3k k_\lambda
\langle\Phi|\sum_{r=1}^2{\alpha^+_r}^a(k){\alpha^-_r}^a(k)|\Phi'\rangle,
\end{equation*}
$|\Phi\rangle$, $|\Phi'\rangle$ $\in$ $\mathscr{H}'$.

In the gauge $\xi=\mu$ the free propagator $\langle0|
T(A^b_\mu(x)A^d_\nu(y))  |0\rangle$ reads
\begin{equation}
\label{136}
\fl (1/i)\mathscr{D}^{bd}_{\mu\nu}(x-y)\equiv\langle0|
T(A^b_\mu(x)A^d_\nu(y))
|0\rangle=-i\mu(g_{\mu\nu}-\frac{\beta}{1+\beta}V_\mu
V_\nu)\delta^{bd}F_c(x-y),
\end{equation}
where
\begin{equation*}
F_c(x-y)=\int\frac{d^4k}{(2\pi)^4}\frac{e^{-ik(x-y)}}{Q(k)+i\epsilon}.
\end{equation*}

As a short example of applications of the theory we shall consider two phenomenas 
that are possible to exist in reactions at high energies. 

For the specific case $G=SU(3)$ and quarks as fermions, the $\beta/(1+\beta)$ term 
in the gluon propagator (\ref{136}) modifies, e.g. the cross-sections for quark-quark 
scattering. For instance, the expression for elastic scattering of two unpolarized 
quarks of mass $m$ and initial (final) momenta $p,k, (p',k')$ in the Born approximation is

\begin{eqnarray*}
\frac{d\sigma}{d\Omega}&=&\frac{2\alpha^2_s}{s}\Biggl[
\frac{F_0}{Q^2(p-p')}+\frac{H_0}{Q^2(p-k')}-\frac{2G_0}{3Q(p-p')Q(p-k')}
\Biggr]-\\
&&-\frac{\beta}{1+\beta}\frac{2\alpha^2_s}{s}\Biggl[
\frac{F_1}{Q^2(p-p')}+\frac{H_1}{Q^2(p-k')}-\frac{2G_1}{3Q(p-p')Q(p-k')}
\Biggr]+\\
&&+\frac{\beta^2}{(1+\beta)^2}\frac{2\alpha^2_s}{s}\Biggl[
\frac{F_2}{Q^2(p-p')}+\frac{H_2}{Q^2(p-k')}-\frac{2G_2}{3Q(p-p')Q(p-k')}
\Biggr],
\end{eqnarray*}

\noindent where we set the permeability $\mu=1$ and

\begin{eqnarray*}
F_0 &=& 16s^2+16st+8t^2-64m^2s+64m^4,\\
H_0 &=& 8s^2+8t^2-64m^2s-64m^2t+192m^4,\\
G_0 &=& -8s^2+64sm^2-96m^4,\\
\\
F_1 &=& 32 \Biggl\{(pV)\biggl[(kV)(p'k')+(k'V)(p'k)\biggr]
+(p'V)\biggl[(k'V)(pk)+(kV)(pk')\biggr]-\\
&&-2(pV)(p'V)(kk')-2(kV)(k'V)(pp')\Biggr\}+32(pp')(kk')+\\
&&+32m^2\biggl[2(pV)(p'V)+2(kV)(k'V)-(pp')-(kk')\biggr]+32m^4,\\
H_1 &=& 32 \Biggl\{(pV)\biggl[(kV)(p'k')+(p'V)(kk')\biggr]
+(k'V)\biggl[(p'V)(pk)+(kV)(pp')\biggr]-\\
&&-2(pV)(k'V)(kp')-2(kV)(p'V)(pk')\Biggr\}+32(pk')(kp')+\\
&&+32m^2\biggl[2(pV)(k'V)+2(kV)(p'V)-(pk')-(kp')\biggr]+32m^4,\\
G_1 &=& 16m^2\biggl[(pV)(kV)+(p'V)(k'V)\biggr]+4\biggl[(p+k)(p'+k')-2m^2\biggr]-\\
&&-8(pk)(p'k'),\\
\\
F_2 &=& 64(pV)(p'V)(kV)(k'V)-32(pV)(p'V)(kk')-32(kV)(k'V)(pp')+\\
&&+32m^2\biggl[(pV)(p'V)+(kV)(k'V)\biggr]-16m^2\biggl[(pp')+(kk')\biggr]+\\
&&+16(pp')(kk')+16m^4,\\
H_2 &=& 64(pV)(p'V)(kV)(k'V)-32(pV)(k'V)(kp')-32(kV)(p'V)(pk')+\\
&& +32m^2\biggl[(pV)(k'V)+(kV)(p'V)\biggr]-16m^2\biggl[(pk')+(kp')\biggr]+16(pk')(kp')+\\
&&+16m^4,\\
G_2 &=& 32(pV)(p'V)(kV)(k'V)-8(pV)\biggl[(p'V)(kk')+(k'V)(p'k)\biggr]-\\
&&-8(kV)\biggl[(k'V)(pp')+(p'V)(pk')\biggr]+4(pp')(kk')-4(pk)(p'k')+\\
&&+4(pk')(p'k).
\end{eqnarray*}

The first line in the expression for the cross-section is (for $n=1$) the known
result from QCD in the Born approximation (see e.g. \cite{27}). 

Another change comes 
from the term $\beta(k\cdot V)^2$ in each denominator
$Q(k)=k^2+\beta(k \cdot V)^2$. For very large $\beta=n^2-1$, this term may lead
to an effective decreasing of the coupling strength.

As a second example we consider the in-medium Yang-Mills equations  
to describe the gluon field emitted by a quark moving in a medium. If $j_\mu^a(x)$ is a 
non-zero quark current in the r.h.s. in (\ref{71}), the corresponding gauge potential 
$A_\mu(x)$ satisfies the equation (in the gauge $\xi=\mu$)

\[
\Bigl(\opensquare + \beta(V \cdot \partial)^2\Bigr)A_\mu^a(x) = j_\mu^a(x)
\]

\noindent for a medium with "refraction" coefficient $n=\sqrt{\beta^2+1}$. In
the $V$-rest frame the equation becomes

\[
\Bigl(n^2\frac{\partial^2}{\partial x_0^2} - \triangle\Bigr)A_\mu^a(x)
=j_\mu^a(x)
\]

\noindent and for a "rapid" quark, a Cherenkov-type gluon radiation will appear.
Thus the description presented here provides a natural framework for the Dremin's
approach aimed at explaining events (such as ring-like structures) which occur 
in hadron matter at high energies.
\section{Summary}
In this paper the in-medium Yang-Mills equations are derived for
a gauge group $G$ in the case of a linear 
response to an external gauge field. As in classical electromagnetism, there
appears naturally a conductivity tensor accompanied, however, by additional
{\it non-Abelian} conductivity tensor and pseudotensor originating from the non-commutative
character of the gauge group $G$. Correspondingly, the medium is characterized by "electric"
and "magnetic" polarizations as well as by a new type of {\it non-Abelian} "electric" and
"magnetic" polarizations. When the latter are assumed to be linearly related with
the fields $E^a$ and $B^a$, the in-medium equations refer to the generalized "electric"
and "magnetic" inductions and for fields  $E'^a$ and $B'^a$ modified by the appearing
susceptibility tensors. If, however, these {\it non-Abelian} polarizations depend non-linearly
on $E^a$ and $B^a$, the restriction to a linear response reduces the modified fields to
$E^a$ and $B^a$, and the in-medium Yang-Mills equations become similar to the macroscopic Maxwell
equations.
In an approximation up to second order in the gauge coupling constant $g$,
the in-medium equations resemble the macroscopic Maxwell equations for
$dim (G)$ copies of the electromagnetic field. In this case the medium
characteristics - the conductivity tensor $\sigma$, the electric
permittivity tensor $\epsilon$ and the magnetic permeability
tensor $\mu$ are expressed in terms of retarded Green's functions
in a way analogous to the case of the in-medium electrodynamics.
The canonical quantization is performed for constant and diagonal
$\epsilon$ and $\mu$ in order to simplify the Euler-Lagrange
equations. A set of $dim (G)$ scalar auxiliary fields $B^a$ is
introduced to avoid working with constrained systems. The
commutation relations among the gauge potentials $A^a_\mu$ and
the field $B^a$ are derived in two ways for a family of
Fermi-like gauges. The first one is straightforward but lengthy
and employs an integral representation of the solutions of equation (\ref{94}),
while the second one is direct and based on a simple
transformation of the both coordinates and fields. For the gauge
choice $\xi=\mu$ we write down the "gluon" propagator
$\mathscr{D}^{bd}_{\mu\nu}$ and evaluate the cross-section for elastic
scattering of two identical unpolarized quarks in the Born approximation. 
A feature of the propagator is the appearance of the term $\beta (k.V)^2$ which 
may lead to an effective decrease of the coupling strength for sufficiently 
large "refraction" coefficient. It is pointed out that the in-medium Yang-Mills equations
\underline{}allow to match Dremin's description of the above mentioned ring-like structures. \\
\\
\textbf {References} \\

\end{document}